\documentclass[conference]{IEEEtran}
\usepackage{graphicx}
\usepackage{amsmath}
\usepackage{amsfonts}
\usepackage{amssymb}
\usepackage{psfrag}
\usepackage{cite}
\usepackage{xcolor}
\usepackage{multirow}
\usepackage{subfigure}
\usepackage[linesnumbered,lined,commentsnumbered]{algorithm2e}
\newtheorem{theorem}{\underline{Theorem}}
\newtheorem{lemma}{\underline{Lemma}}

\newtheorem{remark}{\underline{Remark}}

\newtheorem{proposition}{\underline{Proposition}}
\newtheorem{definition}{\underline{Definition}}

\newcommand{\QED}{{\rm $\blacksquare$}}

\begin{document}
\title{Noncoherent and Non-orthogonal Massive SIMO for Critical Industrial IoT Communications}
\author{He Chen, Zheng Dong, and Branka Vucetic
\\
The University of Sydney, Australia,
Email: \{he.chen, zheng.dong, branka.vucetic\}@sydney.edu.au
}

\maketitle

\begin{abstract}
Towards the realization of ultra-reliable low-latency wireless communications required in critical industrial Internet of Things (IIoT) applications, this paper presents a new noncoherent and non-orthogonal massive single-input multiple-output (SIMO) framework, in which a large number of antennas are deployed to provide high spatial diversity so as to enable ultra-reliable communications, and noncoherent transmission and non-orthogonal multiple access techniques are applied to effectively reduce the latency at the physical and data link layers, respectively. A two-user IIoT system is considered to elaborate the key design principle of our framework, in which two controlled nodes (CNs) transmit their data to a common managing node (MN) on the same time-frequency resource block, and the MN implements the noncoherent maximum likelihood detector to recover the transmitted symbols of both CNs from the received sum signal. We analyze the error performance of the considered system and then minimize the system error probability by jointly designing the constellations of both CNs. Simulation results show that our design has lower error probability than existing designs.
\end{abstract}


\section{Introduction}
The Internet of Things (IoT), aiming to create a smart world by connecting everyday objects and surrounding environments to the Internet, is expected to pervade all aspects of our daily lives and fundamentally alter the way we interact with our physical environment \cite{Al-Fuqaha2015IoT}. The applications of IoT in industrial sectors, termed the Industrial IoT (IIoT) or Industrial Internet, has attracted tremendous attention from governments, academia and industry, for its substantial potential to transform various industry verticals such as electricity, transportation, healthcare, and manufacturing \cite{Xu2014IoT, GEIIOT}. As defined by General Electric (GE), the IIoT refers to ``the network of a multitude of industrial devices connected by communications technologies that results in systems that can monitor, collect, exchange, analyze, and deliver valuable new insights like never before" \cite{GEIIOT}. From this definition, we note that communication technologies play a critical role in realizing the vision of the IIoT.

Critical industrial use cases normally involve real-time closed-loop control, where a failure of communication may lead to serious economic losses and safety accidents \cite{Huang2018New}. Such applications pose stringent performance requirements on the industrial communication networks, with high reliability of packet error rate down to $10^{-9}$ and ultra-low latency at the level of sub-microsecond \cite{Chen2018mag}. These strict requirements are far beyond what latest wireless technologies can provide, and thus have been satisfied by applying wired network infrastructure \cite{Huang2018New}. Nevertheless, wireless communications have several benefits over the currently-used wired infrastructure: low deployment and maintenance cost, easier deployment in scenarios where cables are difficult to deploy, and high long-term reliability by avoiding the wear and tear issues \cite{Luvisotto2017ultra}. There is an emerging consensus that developing ultra-reliable low-latency (URLL) wireless is essential to fully unlock the potential of the IIoT.

In wireless communications, diversity techniques have been used as the main measures to boost system reliability \cite{tse2005fundamentals}. Among various diversity techniques, spatial diversity which is achieved by equipping the transmitter and/or the receiver multiple antennas, is particularly appealing for realizing URLL wireless since it does not need extra resources in time or spectrum domain for high reliability. Considering the ultra-high reliability required by critical IIoT use cases, deploying a massive number of antennas at the transmitter and/or the receiver has been regarded as one of the most promising technologies for URLL wireless \cite{Popovski2018Wireless}. This technology is generally referred to as massive multiple-input multiple-output (MIMO). In this paper, we term it massive single-input multiple-output (SIMO) when only a single antenna is equipped at the transmitter side.

On the other hand, achieving low latency down to the sub-millisecond level in wireless communications is highly challenging. This involve a departure from the underlying theoretical principles of wireless communications---Today's wireless communication networks have been built to maximize data rates and network capacity with latency suited to human perception (i.e., at the level of tens of milliseconds) \cite{Chen2018mag}. Realizing this several orders of magnitude reduction will require significant latency deduction from various layers of the protocol stacks. Industrial networks are typically based on reduced protocol stacks. As such, reducing the latency of the physical and the data link layers is of great importance \cite{Vitturi_IEM_2013}. At the physical layer, considering the fact the data packet (e.g., a sensor data or a control command) in industrial networks is generally very short, shortening the physical layer overheads is an effective method to reduce the latency. There has recently been a line of research focusing on the design of noncoherent single-user massive SIMO systems so as to reduce the channel estimation overhead at the physical layer \cite{Manolakos2016Energy,jing2016design,Popovski2018Wireless,Gao_IoT_2018}, in which different modulators and detectors were designed and analyzed, and time-division multiple access (TDMA) was implicitly assumed to be adopted at the data link layer.

At the data link layer, an effective measures to achieve low latency is to implement non-orthogonal multiple access (NOMA) to replace the currently-used orthogonal TDMA. In NOMA, multiple transmitters are allowed to transmit simultaneously on the same time-frequency block so as to reduce the cycle time, which is defined as the minimum time needed for all the controlled nodes (CNs) to communicate to their managing node (MN) once, and has been the widely-used latency measure for industrial control systems \cite{Luvisotto2017ultra}. To this end, references \cite{Chowdhury2016Scaling} and \cite{zhang2018physically} have recently proposed to jointly optimize the modulation constellations of multiple users in massive SIMO systems to ensure that the symbols transmitted by multiple users at the same time are as distinguishable as possible at the receiver side. In these designs, the minimum Euclidean distance (MED) design criterion was adopted, which aims to maximize the minimum distance between signal points on the sum (composite) constellation at the receiver side. However, as shown in our previous work \cite{Gao_IoT_2018}, for the single-user case, the MED design criterion is obviously suboptimal in terms of system error performance, and adopting the MED design criterion may lead to considerable performance loss. To achieve higher reliability, in \cite{Gao_IoT_2018} we developed a symbol-error-rate-minimization (SERM) design criterion for single-user noncoherent massive SIMO systems. Nevertheless, how to extend the SERM design criterion to address the constellation design for noncoherent \emph{multiuser} massive SIMO with NOMA is, to the best knowledge of the authors, still an open problem in the literature.

As the first effort to fill the aforementioned gap, in this paper we consider the constellation design problem for two-user noncoherent and non-orthogonal massive SIMO systems using the SERM design criterion, in which two single-antenna CNs transmit to a common MN equipped with a large number of antennas at the same time. In doing this, we constrain our design to the case where the constellations of the two CNs are superimposed in a nested manner at the MN side. The MN adopts the optimal noncoherent maximum likelihood (ML) detector to decode the transmitted symbols of both CNs from the received sum signal. We derive closed-form expression for the system SER (SSER) of the considered system, which is defined as the probability that the symbols transmitted by the two CNs are not both decoded correctly. We formulate an SSER minimization problem to jointly optimize the constellations of the two CNs, while subject to their individual average power constraints. The formulated problem is a complex multi-ratio fractional programming (FP) problem, which is in general NP-hard and thus is difficult to resolve \cite{Shen_TSP_2018}. Motivated by this, we simplify the problem to a max-min FP problem by resorting to an asymptotic analysis for the regime that the number of antennas at the MN goes to infinity. We resolve the simplified problem and attain its optimal solution in closed-form, which serves as the asymptotically optimal solution to the original problem. Simulation results are provided to demonstrate that our design is superior to the existing designs adopting the MED design criterion.

\section{System Model}
Consider the uplink scenario of a wireless IIoT system, where two\footnote{Note that the considered system can consist of multiple two-CN pairs, which access the wireless medium in an orthogonal manner.} single-antenna controlled nodes (CNs) transmit their data (e.g., status information) to a managing node (MN), which is the central controller unit of the system and is equipped with $N$ ($N \gg 2$) antennas. To reduce the system circle time, the two CNs are allowed to transmit simultaneously to the MN on the same time-frequency resource block. By employing a discrete-time complex baseband-equivalent model, the received signal vector $\mathbf{y}=[y_1, y_2, \ldots, y_N]^T$ at the MN can be written as
\begin{align}\label{eqn:vecrecsignal}
\mathbf{y}=\mathbf{H}\mathbf{x} +{\boldsymbol\xi},
\end{align}
where $\mathbf{x}=[ x_1, x_2]^T$ represents the transmitted signal vector with $x_k$, $k = 1,2$, denoting the transmitted symbol of the $k$-th CN \emph{equiprobably} drawn from the respective constellation ${\mathcal X}_k$, ${\boldsymbol\xi}$ is the circularly-symmetric complex Gaussian (CSCG) noise vector with covariance $\sigma^2 \mathbf{I}_N$, and $\mathbf{H}=\mathbf{G}\mathbf{D}^{1/2}$ denotes the $N\times 2$ complex channel matrix between the two CNs and $N$ receiving antennas at the MN. We assume that all the entries of $\mathbf{G}$ are i.i.d. CSCG distributed with unit variance to characterize the local scattering fading, $\mathbf{D} ={\rm diag}\{\beta_1, \beta_2\}$ ($\beta_k>0$) is a diagonal matrix which captures the large-scale propagation loss due to the distance and shadowing effect. We also let $\mathbf{h}_n=[h_{1,n}, h_{2,n}]^T$ denote the $n$-th column of $\mathbf{H}$. 
To further reduce the system cycle time, we assume that no instantaneous channel estimation is performed. As such, $\mathbf{G}$ is completely unknown and the noncoherent detection is adopted at the MN to recover the transmitted signals from the two CNs. Nevertheless, the matrix $\mathbf{D}$ is assumed to be available at the MN since it changes much slower and thus can be estimated with much lower overhead compared with the estimation of instantaneous channel coefficients \cite{WagnerTIT12}.

\subsection{Noncoherent Maximum-Likelihood Detector}
For the considered noncoherent multiuser SIMO system with uniform inputs, it is known that the noncoherent ML decoder is optimal in the sense that it minimizes average probability of error of the received sum signal at the MN \cite{key1993fundamentals}. To proceed, we note that~\eqref{eqn:vecrecsignal} can be rewritten as $\mathbf{y}= \mathbf{G}\mathbf{D}^{1/2}\mathbf{x} +{\boldsymbol \xi}$. As all the entries of $\mathbf{G}$ and $\boldsymbol \xi$ are i.i.d. Gaussian, we immediately have $\mathbb E[\mathbf{y}] =\mathbf{0}$. By noting $\mathbf{y}^T= \mathbf{x}^T\mathbf{D}^{1/2}\mathbf{G}^T +{\boldsymbol \xi}^T$, and with the help of~\cite{Petersen12}, we have
\begin{align}
\mathbf{y}={\rm vec} (\mathbf{y}^{T})=(\mathbf{I}_N\otimes \mathbf{x}^T\mathbf{D}^{1/2}){\rm vec}(\mathbf{G}^T)+{\boldsymbol \xi}.
\end{align}
Then, the covariance matrix of $\mathbf{y}$ can be given by
\begin{align*}
\mathbf{R}_{\mathbf{y}|\mathbf{x}}
&=\mathbb{E} \{\mathbf{y}\mathbf{y}^H\}=\mathbb{E} \Big\{\big [\big (\mathbf{I}_N\otimes \mathbf{x}^T\mathbf{D}^{1/2}){\rm vec}(\mathbf{G}^T)+{\boldsymbol \xi}\big ]\\
&~~~~~~~~~~~~~~~~~~~~~\big [(\mathbf{I}_N\otimes \mathbf{x}^T\mathbf{D}^{1/2}){\rm vec}(\mathbf{G}^T)+{\boldsymbol \xi}\big ]^H\Big\}\\
&=(\mathbf{x}^T \mathbf{D}\mathbf{x}^* +\sigma^2)\mathbf{I}_N=(\mathbf{x}^H \mathbf{D}\mathbf{x} +\sigma^2)\mathbf{I}_N =c({\bf x})\mathbf{I}_N,
\end{align*}
where $c({\bf x})$ is the sufficient statistic of the input signal, which is defined as
\begin{align}\label{eqn:c_def}
c({\bf x}) =\mathbf{x}^H \mathbf{D}\mathbf{x} +\sigma^2 = \sum\nolimits_{k=1}^2 {\beta_k |x_k|^2}  +\sigma^2.
\end{align}
The probability density function (PDF) of the received signal $\mathbf{y}$ at the MN conditioned on the input signal $\mathbf{x}$ can thus be given by
\begin{align}\label{eqn:conditionalpdf}
f({\mathbf{y}|\mathbf{x}})=\frac{1}{\pi^N c^N({\bf x})} \exp\Big (-\frac{\|\mathbf{y}\|^2}{c({\bf x})}\Big ).
\end{align}
The noncoherent ML detector aims to estimate the transmitted information by carrying out the following optimization problem:
\begin{align}\label{eqn:MLjointdector}
\hat{\mathbf{x}}={\arg\min}_{\mathbf{x}}\ln f({\mathbf{y}|\mathbf{x}}).
\end{align}
Combining~\eqref{eqn:conditionalpdf} and~\eqref{eqn:MLjointdector}, we have
\begin{align}\label{eqn:MLjointdector_1}
\hat{\mathbf{x}}={\arg\min}_{\mathbf{x}}~\frac{\|\mathbf{y}\|^2}{c({\bf x})} + N \ln c({\bf x}).
\end{align}
We can observe from \eqref{eqn:MLjointdector_1} that the phase information of the input signal is lost. As such, we can only modulate the information to be transmitted on the power of the transmitted signal (i.e., $|x_k|^2$) in the considered system, which is termed energy-based modulation in \cite{Manolakos2016Energy,zhang2018physically}. Note that we hereafter use energy and power interchangeably as the symbol duration of the considered system is fixed. We define the (nonnegative) constellation of each CN as a collection of the power of the transmitted symbols. For notation simplicity, we assume that both CNs use the same $M$-ary constellation\footnote{It is worth mentioning that our design framework can be extended to the case with all CNs using distinct orders of modulation, where a more complicated notation system is required.}. We then use $\mathcal{X}_k=\{s_{k,i}\}_{i=1}^{M}$ to denote the constellation of the $k$-th CN,  $k=1,2$, or equivalently $|x_k|^2 \in \mathcal{X}_k$. We assume that each CN is subject to an individual average power constraint given by
\begin{align}\label{eqn:power_constraint}
{\sum\nolimits_{i=1}^{M} s_{k,i}}/{M}\le {P_k}, \quad k=1,2,
\end{align}
where $P_k$ is the average power constraint of the $k$-th CN. For the sake of notation later, we further define the constellation set $\mathcal{A}_k=\{a_{k,i}\}_{i=1}^{M}=\{\beta_k s_{k,i}\}_{i=1}^{M}$. The power constraint in \eqref{eqn:power_constraint} is then equivalent to
\begin{align}\label{eqn:power_constraint1}
{\sum\nolimits_{i=1}^{M} a_{k,i}}/{M}\le \beta_k {P_k}, \quad k=1,2.
\end{align}
Motivated by the fact that uniform constellations is preferred in most practical communication systems, we consider that all ${\mathcal A}_k$'s are uniform constellations. We then can express the constellation set $\mathcal{A}_k$ of the $k$-th CN as $\mathcal{A}_k=\{m \bar {\delta}_k + q_k\}_{m=0}^{M-1}$.
The individual average power constraint can be simplified as
\begin{align}\label{eqn:power_constraint2}
q_k + \frac{M-1}{2} \bar{\delta}_k \le \beta_k {P_k}, \quad k=1,2.
\end{align}
Without loss of generality, we assume that $\beta_1 P_1 \le \beta_2 P_2 $. We then can set $\bar{\delta}_1 \le \bar{\delta}_2$.

As we can see from \eqref{eqn:conditionalpdf}, the PDF of the received signal conditioned on the input signal, $f({\mathbf{y}|\mathbf{x}})$, is completely characterized by the sufficient channel statistic function $c(\bf x)$. Furthermore,
$c(\bf x)$ involves the summation of elements drawn from the sets $\mathcal{A}_k$, $k=1,2$. To formally model this, we define the sum constellation $\mathcal{B} =\left\{ \sum_{k=1}^2 a_k : a_k \in {\mathcal A}_k \right\} $. To ensure that in the noise-free case, the receiver can always distinguish all the transmitted symbols once any sum signal $b$, $ \forall~b \in {\mathcal B}$, is received, we require that the set ${\mathcal B}$ must be \emph{uniquely factorable} \cite{dong16jstsp}, which is denoted by ${\mathcal B}= \mathcal{A}_1 \uplus \mathcal{A}_2$ and is formally defined as:
\begin{definition}\label{}
The set ${\mathcal B}$ is \emph{uniquely factorable} if and only if $|\mathcal{B}|=\prod_{k=1}^{2} |\mathcal{A}_k| = M^2$ . That is, for $b =\sum_{k=1}^2 a_k$ and $b'=\sum_{k=1}^2 a_k'$, the equality $b=b'$ is equivalent to $(a_1, a_2)=(a_1',a_2')$.~\hfill\QED
\end{definition}

\begin{figure}
  \centering
  \includegraphics[width=0.8\linewidth]{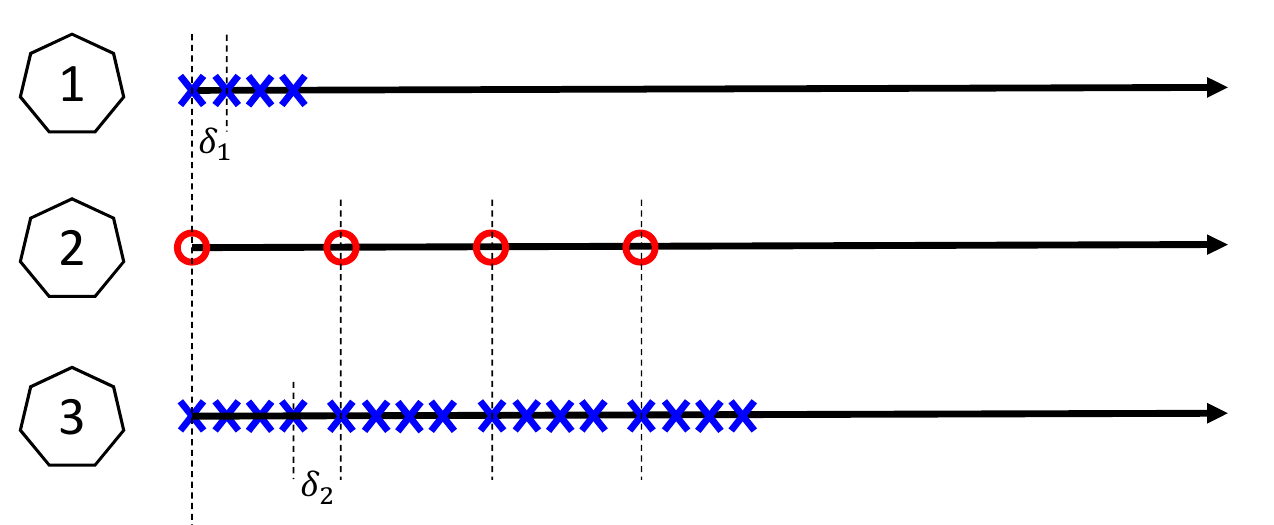}
  \caption{Illustration of the nested sum constellation of two nonnegative uniform constellations of order 4 with $q_1 = q_2 = 0$, where the sum of the first and second constellations produces the third constellation. We also note that the sum constellation is uniquely determined by the three distances $\delta_1$ and $\delta_2$.}\label{Fig:sum constellation}
\end{figure}

In other words, we require the term $c({\bf x})$ defined in \eqref{eqn:c_def} to have a one-to-one correspondence with the transmitted signal vector $\bf x$. Then, the transmitted signal of each CN can be uniquely determined if the sum signal can be correctly detected. With the aid of the uniquely factorable property between the sum constellation and the separate constellation used by each CN, the optimization problem \eqref{eqn:MLjointdector_1} to be solved by the noncoherent ML detector can be simplified into the detection of the received sum signal as:
\begin{align}\label{eqn:MLjointdector2}
\hat{c}=\mathop{\arg\min}\limits_{c \in {\mathcal C}}~\frac{\|\mathbf{y}\|^2}{c} + N \ln c,
\end{align}
where ${\mathcal C} = \{c_\ell \}_{\ell = 1}^{M^2} = \{b_\ell + \sigma^2\}_{\ell = 1}^{M^2}$.

As an initial effort, in this paper we constrain our design framework to the scenario where the signal constellations of the two CNs are superimposed in a \emph{nested} manner over the air. That is, the distance between the two end points of the smaller constellation is less than the distance between the adjacent points of the larger constellation. Mathematically, we have $\bar{\delta}_{2} > (M-1) \bar{\delta}_{1}$. To facilitate the understanding, we illustrate the process of a nested summation of two nonnegative uniform constellations of the same order 4 in Fig. \ref{Fig:sum constellation}. We can see from this figure that the nested summation of the constellations significantly reduce the minimum Euclidean distance of the sum constellation at the receiver side. Fortunately, the resultant performance loss can be effectively compensated by the large number of antennas equipped at the MN.

We observe from Fig. \ref{Fig:sum constellation} that we can define a new notation $\delta_2$, as the difference between the minimum Euclidean distance of constellation ${\mathcal A}_2$ (i.e., $\bar{\delta}_2$) and the Euclidean distance of the two end points on the constellation ${\mathcal A}_1$. We also let $\bar{\delta}_1 = {\delta}_1 $. We will later show that using $\delta_k$ instead of $\bar{\delta}_k$ can simplify the presentation of the optimization problem. With this new definition, the resultant sum constellation of both CNs can be completely characterized by $\left\{\delta_k\right\}_{k=1}^2$ and $\left\{q_k\right\}_{k=1}^2$. Specifically, $\left\{b_\ell\right\}_{\ell=1}^{M^2}$ and $\left\{c_\ell\right\}_{\ell=1}^{M^2}$ are both nonnegative weighted sum of $\left\{\delta_k\right\}_{k=1}^2$ and $\left\{q_k\right\}_{k=1}^2$. That is, given the modulation size $M$, $\left\{\delta_k\right\}_{k=1}^2$ and $\left\{q_k\right\}_{k=1}^2$, we can readily enumerate the expressions of $\left\{c_\ell\right\}_{\ell=1}^{M^2}$. In the meanwhile, the constellations of the two CNs, ${\mathcal A}_1$ and ${\mathcal A}_2$, can also be determined. Hereafter, $\left\{\delta_k\right\}_{k=1}^2$ and $\left\{q_k\right\}_{k=1}^2$ are the key parameters to be optimized in this paper. Furthermore, by applying the mathematical induction, the average power constraints of both CNs given in \eqref{eqn:power_constraint2} can be further expanded as
\begin{align}
&q_1 + \frac{M-1}{2} {\delta}_1 \le \beta_1 {P_1}, \label{eqn:rho_constraint1}\\
& q_1+q_2 + \frac{M-1}{2} \left[ \left( {M - 1} \right){\delta _1} +  {\delta _2} \right] \le \beta_2 {P_2}. \label{eqn:rho_constraint2}
\end{align}

\section{Error Performance Analysis and Problem Formulation}
\subsection{Optimal Decision Regions and Error Performance}
We subsequently derive the optimal decision regions of $\|\mathbf{y}\|^2$ in the non-coherent ML detector for a given group of constellations $\{{\mathcal A}_k \}_{k=1}^2$ (i.e., the set $\mathcal C$ is given). Without loss of generality, we consider that all the elements of the set $\mathcal C$ are arranged in an ascending order such that $c_{\ell} < c_{\ell + 1}$ for $\ell = 1,2,\ldots,M^2 - 1$. We now resolve the optimization problem of the adopted noncoherent ML detector given in \eqref{eqn:MLjointdector2} and attain the following theorem on the optimal decision regions of $\|\mathbf{y}\|^2$:
\begin{theorem}\label{thm:decisionrule}
The optimal decision regions of $\|\mathbf{y}\|^2$ for the adopted non-coherent ML detector can be written as
\begin{align}\label{eqn:decisionrule}
	\hat{c}=\begin{cases}
	c_1, &{\rm if~} \frac{\|\mathbf y\|^2}{N}\leq d_1;\\
	c_\ell,&{\rm if~} d_{\ell-1}<\frac{\|\mathbf y\|^2}{N}\leq d_\ell,~\ell=2,\ldots, M^2-1;\\
	c_{|\mathcal B|},& {\rm if~} \frac{\|\mathbf y\|^2}{N}>d_{M^2-1},
	\end{cases}
	\end{align}
where $d_\ell = c_{\ell+1}\mu \left(\frac{c_{\ell + 1}}{c_{\ell}}\right)$ with $\mu(x) = \frac{\ln x}{x-1}$.~\hfill\QED
\end{theorem}
The proof is omitted due to space limitation.

\begin{remark}
In Theorem~\ref{thm:decisionrule}, we have simplified the noncoherent ML detector into an average received power-based detector. Specifically, the MN only needs to the estimate the average power of the received signal (i.e., $\frac{\|\mathbf y\|^2}{N}$) to detect the sum signal $c$. Then, the respective signal transmitted by each CN can be uniquely determined by using the one-to-one correspondence between $c$ and $\bf x$. ~\hfill\QED
\end{remark}

We now analyze the successful transmission probability of the signal vector ${\bf x}_{\ell}$. Recall that ${\bf x}_{\ell}$ and $c_{\ell}$ have one-to-one correspondence. Denote by ${{{{\left\| {{\bf{y}}\left( {{{\bf{x}}_{\ell}}} \right)} \right\|}^2}}}$ the received signal at the MN when ${\bf x}_\ell$ is transmitted by the CNs. According to Theorem \ref{thm:decisionrule}, the successful transmission probability of the signal vector ${\bf x}_{\ell}$, denoted by $P_{c,\ell}$, can be written as
\begin{align}\label{eqn:PEP}
	P_{c,\ell}=\begin{cases}
	\Pr \left( {\frac{{{{\left\| {{\bf{y}}\left( {{{\bf{x}}_1}} \right)} \right\|}^2}}}{N} \le {d_1}} \right), &{\rm if~}\ell = 1;\\
	\Pr \left( {{d_{\ell  - 1}} < \frac{{{{\left\| {{\bf{y}}\left( {{{\bf{x}}_{\ell}}} \right)} \right\|}^2}}}{N} \le {d_\ell }} \right),&{\rm if~} 2\le \ell\le M^2-1;\\
	\Pr \left( \frac{{{{\left\| {{\bf{y}}\left( {{{\bf{x}}_{M^2}}} \right)} \right\|}^2}}}{N}>d_{M^2-1} \right),& {\rm if~} \ell = {M^2}.
	\end{cases}
	\end{align}
In this paper, we consider the scenario that the MN needs to collect both CNs' information correctly so as to make a further decision. In this case, the MN will claim an error if the sum signal as a whole is decoded erroneously. We define the probability of such an error as the system symbol error rate (SSER). Recall that the transmitted signals of both CNs are drawn from their respective constellations with the same probability. We thus can express the SSER as
\begin{equation}\label{eqn:SSER}
{P_e} = 1 - \frac{1}{{{M^2}}}\sum\limits_{\ell  = 1}^{{M^2}} {P_{c,\ell}}.
\end{equation}

To proceed, we note that the random variable ${\frac{{{{\left\| {{\bf{y}}\left( {{{\bf{x}}_\ell}} \right)} \right\|}^2}}}{c_\ell}}$ follows a Chi-squared distribution and its cumulative distribution function (CDF) is given by
\begin{align}\label{eqn:G_func}
G\left( x \right) = 1 - \exp \left( { - x} \right)\sum\limits_{m = 0}^{N - 1} {\frac{{{x^m}}}{{m!}},} ~x>0.
\end{align}
We can further simplify \eqref{eqn:PEP} as follows
\begin{align}
	P_{c,\ell}&=\begin{cases}
	\Pr \left( {\frac{{{{\left\| {{\bf{y}}\left( {{{\bf{x}}_1}} \right)} \right\|}^2}}}{c_1} \le {N \frac{c_2}{c_1}\mu \left(\frac{c_{2}}{c_{1}}\right)}} \right) ,~\hfill{\rm if~}\ell = 1;\\
	\Pr \left( {N \mu \left(\frac{c_{\ell }}{c_{\ell-1}}\right) < \frac{{{{\left\| {{\bf{y}}\left( {{{\bf{x}}_{\ell}}} \right)} \right\|}^2}}}{c_\ell} \le {N \frac{c_{\ell+1}}{c_\ell}\mu \left(\frac{c_{\ell+1}}{c_{\ell}}\right)} } \right),\\~\hfill{\rm if~} 2\le \ell\le M^2-1;\\
	\Pr \left( \frac{{{{\left\| {{\bf{y}}\left( {{{\bf{x}}_{M^2}}} \right)} \right\|}^2}}}{c_{M^2}}> N \mu \left(\frac{c_{M^2 }}{c_{M^2-1}}\right) \right), ~\hfill {\rm if~} \ell = {M^2}.
	\end{cases}\\
&=\begin{cases}\label{eqn:PEP1}
	 G\left( {N \frac{c_2}{c_1}\mu \left(\frac{c_{2}}{c_{1}}\right)} \right), ~\hfill{\rm if~}\ell = 1;\\
	 G\left( {N \frac{c_{\ell+1}}{c_\ell}\mu \left(\frac{c_{\ell+1}}{c_{\ell}}\right)} \right) - G\left( N \mu \left(\frac{c_{\ell }}{c_{\ell-1}}\right) \right)  ,\\~\hfill{\rm if~} 2\le \ell\le M^2-1;\\
	1- G\left( N \mu \left(\frac{c_{M^2 }}{c_{M^2-1}}\right) \right),~\hfill {\rm if~}  \ell = {M^2}.
	\end{cases}
	\end{align}

Substituting \eqref{eqn:PEP1} into \eqref{eqn:SSER} and making necessary manipulations, we can obtain a closed-form expression for the SSER as follows
\begin{align}
{P_e}
&= \frac{1}{{{M^2}}}\sum\limits_{\ell  = 1}^{{M^2} - 1} {F\left( {\frac{{{c_{\ell  + 1}}}}{{{c_\ell }}}} \right)},
\end{align}
where
\begin{align}\label{eqn:F_func}
F\left( t \right) = 1 + G\left( {N\mu \left( t \right)} \right) - G\left( {Nt\mu \left( t \right)} \right)
\end{align}
is defined for notation simplicity.

\subsection{Problem Formulation}
We are now ready to formulate a SSER minimization problem for the considered system, in wihch we optimize the constellations of both CNs (i.e., $\left\{\delta_k\right\}_{k=1}^2$ and $\left\{q_k\right\}_{k=1}^2$) while considering the individual average power constraint of each CN. Mathematically, we have
\begin{align}
{\rm({\bf P1})} \quad &\mathop {\min }\limits_{\left\{ {{\delta _k}} \right\}_{k = 1}^2, \left\{q_k\right\}_{k=1}^2} {P_e}= \frac{1}{{{M^2}}}\sum\limits_{\ell  = 1}^{{M^2} - 1} {F\left( {\frac{{{c_{\ell  + 1}}}}{{{c_\ell }}}} \right)}, \\
&~~~~~~{\rm{ s}}{\rm{.t}}{\rm{. }} \quad \delta_k \ge 0 ,~ q_k\ge0,~ \eqref{eqn:rho_constraint1}, ~ \eqref{eqn:rho_constraint2},
\end{align}
where we recall that $\left\{c_\ell\right\}_{\ell=1}^{M^2}$ are nonnegative weighted sum of $\left\{\delta_k\right\}_{k=1}^2$ and $\left\{q_k\right\}_{k=1}^2$. We can see that $\bf{(P1)}$ is a multi-ratio fractional programming (FP) problem. More specifically, it is a sum-of-functions-of-ratio problem, which is generally NP-hard \cite{Shen_TSP_2018}.

We now try to simplify $\bf{(P1)}$ by investigating the characteristics of its objective function and constraints. We first arrive at the following lemma regarding the optimal value of $\left\{q_k\right\}_{k=1}^2$:
\begin{lemma}\label{lemma:q_optimal}
The optimal values of $\left\{q_k\right\}_{k=1}^2$ in $\bf{(P1)}$ are $q_1^* =  q_2^* = 0$.
\end{lemma}
The proof is omitted due to space limitation.
\begin{remark}
Lemma~\ref{lemma:q_optimal} indicates that all the optimal constellations must include the origin. This result can be understood intuitively as follows: When not all the constellations used by the CNs include zero, the resultant sum constellation will not include the zero. In this case, we can always move the most left-side constellation point of the sum constellation to the origin to further reduce the SSER without violating the average power constraints of the CNs. As such, all the optimal constellations used by CNs should include the origin.~\hfill\QED
\end{remark}

Applying Lemma~\ref{lemma:q_optimal}, we reduce $\bf{(P1)}$ to the following optimization problem
\begin{align}
&{\rm({\bf P1.1})} \quad \mathop {\min }\limits_{\delta_1, \delta_2} {P_e}= \frac{1}{{{M^2}}}\sum\nolimits_{\ell  = 1}^{{M^2} - 1} {F\left( {\frac{{{c_{\ell  + 1}}}}{{{c_\ell }}}} \right)},\notag \\
&~~~~~~~~~~~~{\rm{ s}}{\rm{.t}}{\rm{. }} ~ \delta_k \ge 0,~
\frac{M-1}{2} {\delta}_1 \le \beta_1 {P_1}, \label{eqn:rho_constraint1_final}\\
&~~~~~~~~~~~~~~~~~  \frac{M-1}{2} \left[ \left( {M - 1} \right){\delta _1} +  {\delta _2} \right] \le \beta_2 {P_2} \label{eqn:rho_constraint2_final}
\end{align}
where $\left\{c_\ell\right\}_{\ell=1}^{M^2}$ are nonnegative weighted sum of $\left\{\delta_k\right\}_{k=1}^2$ \emph{only}. Though we have simplified the original $\bf{(P1)}$ by removing half of the variables to be optimized, the new $\bf{(P1.1)}$ is still difficult to resolve due to the complicated structure of the objective function. To the best knowledge of the authors,
only a stationary point (local optimality) of $\bf{(P1.1)}$ can be efficiently achieved by applying the latest quadratic transform algorithm developed in \cite{Shen_TSP_2018}. Motivated by this issue, in the subsequent section we will study the asymptotic case with the number of antennas at the MN (i.e., $N$) approaching infinity so as to attain the asymptotically optimal solution to $\bf{(P1.1)}$, i.e., asymptotically optimal constellation design for the considered IIoT system.

\section{Asymptotically Optimal Constellation Design}
In this section, we perform the asymptotic analysis of the SSER for the regime that the number of antennas equipped at the MN goes to infinity (i.e., $N  \to \infty$), so as to further simplify the objective function of $\bf{(P1.1)}$. In our previous work \cite{Gao_IoT_2018}, we have conducted similar asymptotic analysis for a single-user noncoherent massive SIMO system. By following a similar procedure, we attain that both the upper and lower bounds of the SSER $P_e$ are monotonically decreasing functions of the term $\min \left\{ {\frac{{{c_{\ell  + 1}}}}{{{c_\ell }}}} \right\}_{\ell  = 1}^{{M^2} - 1}$. We omit the detailed derivation for brevity and refer interested readers to \cite[Theorem 3]{Gao_IoT_2018} and its proof for details. It is worth mentioning that the asymptotic expression has shown to be very tight and can approach its exact counterpart when $N$ is moderately large~\cite{Gao_IoT_2018}. The nice feature identified in the asymptotic analysis indicates that minimizing the SSER is equivalent to maximizing the term $\min \left\{ {\frac{{{c_{\ell  + 1}}}}{{{c_\ell }}}} \right\}_{\ell  = 1}^{{M^2} - 1}$. Mathematically, we can simplify $\bf{(P1.1)}$ to the following problem
\begin{align}
{\rm{\bf (P2)}} \quad \mathop {\max }\limits_{\left\{ {{\delta _k}} \right\}_{k = 1}^2} \min \left\{ {\frac{{{c_{\ell  + 1}}}}{{{c_\ell }}}} \right\}_{\ell  = 1}^{{M^2} - 1}, ~{\rm{ s}}{\rm{.t}}{\rm{. }} \quad\eqref{eqn:rho_constraint1_final}, \eqref{eqn:rho_constraint2_final},
\end{align}
which is a max-min-ratio problem. In fact, $\bf{(P2)}$ can be directly resolved by applying the quadratic transform algorithm developed in \cite{Shen_TSP_2018}. After a careful observation at the ratios in the objective function of $\bf{(P2)}$, we find that we can further simplify $\bf{(P2)}$ due to the following two important observations:
\begin{itemize}
  \item \emph{\underline {Observation 1}}: By recalling the definitions of $\left\{c_\ell\right\}_{\ell=1}^{M^2}$ and $\left\{\delta_k\right\}_{k=1}^2$, we notice that for any $\ell$, the difference between $c_{\ell+1} - c_{\ell}$ is always equal to one of the $\delta_k$'s. By this observation, we divide all the $M^2 - 1$ ratios in the objective function of $\bf{(P2)}$ into two groups, with the $k$th group being denoted by $\left\{ {\frac{{{c_\ell } + {\delta _k}}}{{{c_\ell }}}} \right\}_{ {\ell  \in \left\{ {\left. {1, \ldots ,{M^2}} \right|{c_{\ell  + 1}} - {c_\ell } = {\delta _k}} \right\}} }$. Note that the number of ratios in each group can be different.
  \item \emph{\underline {Observation 2}}: For a given $\delta _k$, the larger the $c_\ell$, the smaller the ratio ${\frac{{{c_\ell } + {\delta _k}}}{{{c_\ell }}}}$. As such, the minimal ratio in the $k$th group is achieved when $\ell$ equals to its maximum possible value in the set ${\ell  \in \left\{ {\left. {1, \ldots ,{M^2}} \right|{c_{\ell  + 1}} - {c_\ell } = {\delta _k}} \right\}}$. We denote the maximum possible value of $\ell$ in the $k$th group as $\ell_{\delta_k}$.
\end{itemize}

By the above two important observations, we have successfully reduced the number of ratios in $\bf{(P2)}$ from $M^2 - 1$ to $2$. By applying the mathematical induction method to enumerate the values of $c_{\ell_{\delta_k}}$'s for given $M$, we successfully simplify $\bf{(P2)}$ to the following problem
\begin{align*}
{\rm{\bf (P3)}} \quad&\mathop {\max }\limits_{{\delta _1},{\delta _2}} \min \left\{ \frac{{M\left( {M - 1} \right){\delta _1} + \left( {M - 1} \right){\delta _2} + {\sigma ^2}}}{{M\left( {M - 1} \right){\delta _1} + \left( {M - 1} \right){\delta _2} + {\sigma ^2} - {\delta _1}}}, \right. \notag\\
&~~~~~~~~~~~~~~~\left. \frac{{{{\left( {M - 1} \right)}^2}{\delta _1} + \left( {M - 1} \right){\delta _2} + {\sigma ^2}}}{{{{\left( {M - 1} \right)}^2}{\delta _1} + \left( {M - 1} \right){\delta _2} + {\sigma ^2} - {\delta _2}}} \right\} \notag\\
&~~~{\rm{s}}{\rm{.t}}{\rm{. }}\quad\eqref{eqn:rho_constraint1_final}, \eqref{eqn:rho_constraint2_final}. \notag
\end{align*}

After some mathematical manipulations, we arrive at the following proposition regarding the optimal solution to $\bf{(P3)}$.
\begin{proposition}\label{prop:optimal_solution_2user}
The optimal solution to $\bf{(P3)}$, denoted by $\delta_1^*$ and $\delta_2^*$, is determined in the following two cases:

$\bullet$~If $\tilde \delta _1^\dag \left( {{2\beta_2P_2}} \right) \le
2\beta_1 P_1$, we have
  $\delta_1^* = \tilde \delta _1^\dag \left( {{2\beta_2P_2}} \right)/(M-1)$, and $\delta_2^* = 2\beta_2P_2/(M-1)-\tilde \delta _1^\dag \left( {{2\beta_2P_2}} \right)$.

$\bullet$~If $\tilde \delta _1^\dag \left( {{2\beta_2P_2}} \right) >
2\beta_1 P_1$, we have
$ \delta_1^* =  {{2\beta_1P_1}} /(M-1)$, and
$\footnotesize \delta_2^* = \frac{{ - \left( {\frac{{{\sigma ^2}}}{{M - 1}} + {2\beta_1P_1}} \right) + \sqrt {{{\left( {\frac{{{\sigma ^2}}}{{M - 1}} - {2\beta_1P_1}} \right)}^2} + \frac{{4\left( {{{\left( {{2\beta_1P_1}} \right)}^2} + {2\beta_1P_1}{\sigma ^2} + \frac{{{2\beta_1P_1}{\sigma ^2}}}{{M - 1}}} \right)}}{{M - 1}}} }}{{2}} $.\\
Here, $\small \tilde \delta _1^\dag \left( {{{\tilde \delta }_2}} \right) = \frac{{ - \left( {{{\tilde \delta }_2} + {\sigma ^2} + \frac{{{\sigma ^2}}}{{M - 1}}} \right) + \sqrt {{{\left( {{{\tilde \delta }_2} + {\sigma ^2} + \frac{{{\sigma ^2}}}{{M - 1}}} \right)}^2} + \frac{{4\left( {\tilde \delta _2^2 + {{\tilde \delta }_2}{\sigma ^2}} \right)}}{{M - 1}}} }}{2}$.
\end{proposition}
The proof is omitted due to space limitation.

Till now we have obtained the asymptotically optimal constellations of the two CNs in the considered system.

\section{Simulation Results}
\begin{figure}
  \centering
  \includegraphics[width=1\linewidth]{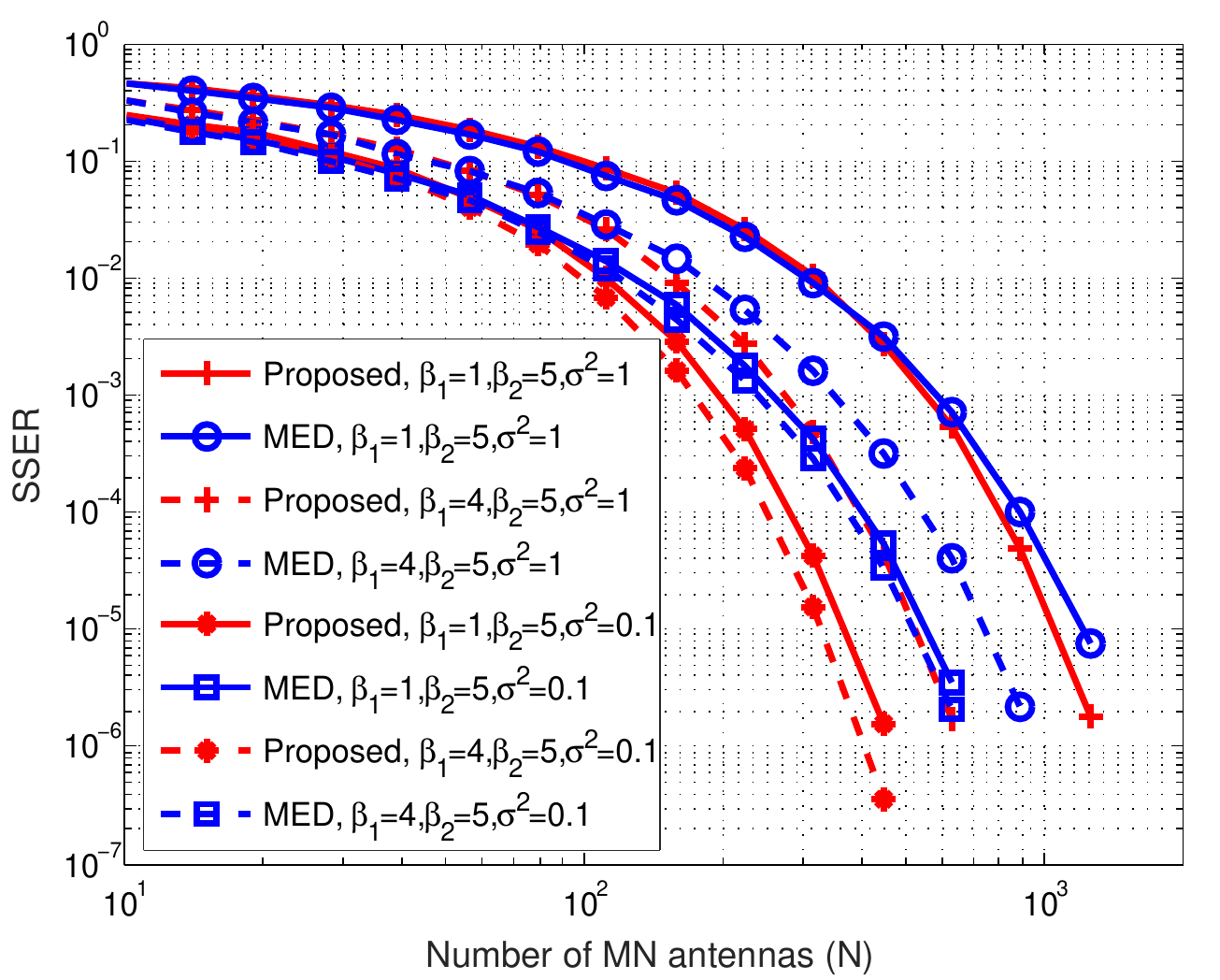}
  \caption{Comparison between our design and the MED design, where $P_1 = P_2 =$ 316\,mW (25\,dBm) and $M = 2$.}\label{Fig:SSER}
\end{figure}

We now present simulation results to compare the SSER performance of the proposed design and the existing design using MED criterion. In doing this, we plot the SSER curves of these two schemes versus the number of antennas equipped at the MN (i.e., $N$) for different values of $\beta_1$, $\beta_2$, and $\sigma^2$ in Fig. \ref{Fig:SSER}. We can see from Fig. \ref{Fig:SSER} that our design is superior to the MED counterpart for all simulated scenarios, as long as $N$ is large enough. Moreover, for a given signal-to-noise ratio (i.e., $\sigma^2$ is fixed), the performance gap of our scheme over the MED one becomes lager when the value of $\beta_1$ is closer to that of $\beta_2$. This is because when $\beta_2$ is fixed, the larger $\beta_1$ gives us more space to optimize the smaller constellation such that the performance gain over the MED design criterion is enlarged.

\section{Conclusions}
In this paper, we developed a new noncoheret and non-orthogonal massive SIMO framework to enable ultra-reliable low-latency wireless needed in emerging critical industrial Internet of Things (IIoT) applications. As the first work within this framework, we have designed a two-user IIoT system, which consists of two single-antenna controlled nodes (CNs) and one managing node (MN) equipped with a large number of antennas. The two CNs transmit their information to the MN simultaneously on the same radio resource, and the MN applies the noncoherent maximum likelihood detector to recover both CNs' information from the received sum signal. We jointly optimized the constellations of both CNs to maximize the system reliability. We managed to find the closed-form expression of the asymptotically optimal solution to the formulated problem. Simulation results demonstrated that the proposed design has better system reliability than the existing designs adopting the minimum Euclidean distance criterion.

As future work, we will extend our framework to arbitrary number of users, and will implement the design on software-defined radio platforms to demonstrate and evaluate its performance in real environments.

\small
\bibliographystyle{ieeetr}
\bibliography{tzzt}

\normalsize


\end{document}